\documentclass[a4paper,11pt]{article}
\pdfoutput=1 % if your are submitting a pdflatex (i.e. if you have
\usepackage{jheppub} % for details on the use of the package, please
\usepackage[T1]{fontenc} % if needed
\notoc

\title{Collaborative Computing Support for Analysis Facilities Exploiting Software as Infrastructure Techniques}

\author[a]{Maria Acosta Flechas,}
\author[b]{Garhan Attebury,}
\author[b]{Kenneth Bloom,}
\author[c]{Brian Bockelman,}
\author[a]{Lindsey Gray,}
\author[a]{Burt Holzman,}
\author[b]{Carl Lundstedt,}
\author[b]{Oksana Shadura,}
\author[a]{Nicholas Smith,}
\author[b]{John Thiltges}

\affiliation[a]{Fermi National Accelerator Laboratory, USA}
\affiliation[b]{University of Nebraska-Lincoln, USA}
\affiliation[c]{Morgridge Institute for Research, USA}

\abstract{Prior to the public release of Kubernetes it was difficult to conduct joint development of elaborate analysis facilities due to the highly non-homogeneous nature of hardware and network topology across compute facilities. However, since the advent of systems like Kubernetes and OpenShift, which provide declarative interfaces for building fault-tolerant and self-healing deployments of networked software, it is possible for multiple institutes to collaborate more effectively since resource details are abstracted away through various forms of hardware and software virtualization. In this whitepaper we will outline the development of two analysis facilities: ``Coffea-casa'' at University of Nebraska Lincoln and the ``Elastic Analysis Facility'' at Fermilab, and how utilizing platform abstraction has improved the development of common software for each of these facilities, and future development plans made possible by this methodology.}

\begin{document} 
\maketitle
\flushbottom

\section{Introduction}

The analysis facilities discussed in this whitepaper were designed from the start to use a container-based infrastructure. Containers provide flexibility, portability and isolation without the additional overhead of virtual machines. Sites that deploy this infrastructure at a wide scale make it easier to add elasticity to the analysis facility; servers for a different purpose (e.g. batch worker nodes) can be reprovisioned on the fly for scheduling analysis tasks. The orchestration tool of choice for containers is Kubernetes. It provides a unified declarative description and configuration language, configuration management, service discovery, service load balancing, automated rollouts and rollbacks, and other features key to providing stable services. 

Kubernetes was originally designed for cloud computing, which adopts a single-tenant model: one user creates and owns an entire cluster. Since its original public release, it has been extended with role-based access controls, policy primitives, and a configurable programmable filter module in front of the API.
Pure Kubernetes is a good fit for facilities such as Coffea-casa which are designed to serve a single experiment.  For multi-tenant facilities there is Red Hat's open-source OKD platform, which is a superset of Kubernetes. OKD incorporates additional security and isolation, adds operations-centric tools, a user-friendly GUI, and additional storage and network orchestration components, making OKD a good choice for the Elastic Analysis Facility. 

\begin{figure}[!ht]
  \centering
  \includegraphics[width=0.7\textwidth]{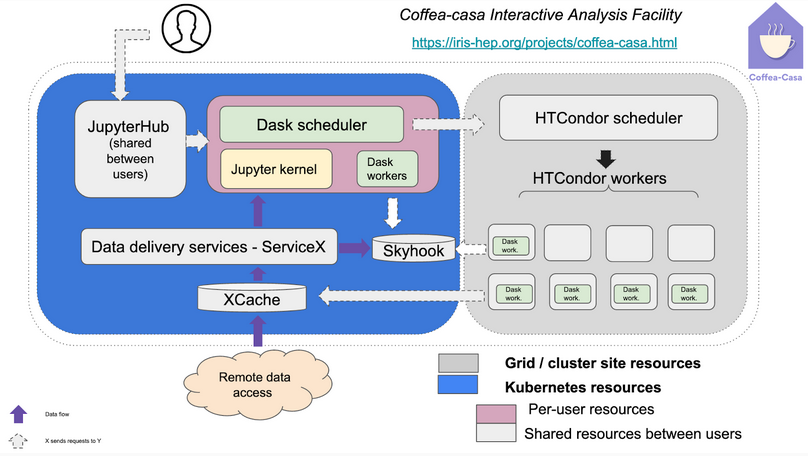}
  \caption{CMS analysis facility at the University of Nebraska.}
  \label{fig1}
\end{figure}

Coffea-casa is an analysis facility prototype built on a combination of Nebraska's multi-tenant Kubernetes resource and their HTCondor pool dedicated for CMS Tier-2 Nebraska computing.  Kubernetes provides a service orchestration layer, allowing the administrative team to programmatically deploy cyberinfrastructure such as custom pods, network connectivity, and block devices.  For example, Kubernetes hosts a JupyterHub  instance that authenticates users and, upon login, provisions a dedicated ``analysis pod'' for the user.  In this case a Kubernetes ``pod'' is the atomic unit of compute and service management (potentially consisting of several containers).  For Coffea-casa, the analysis pod consists of a Jupyter notebook, a Dask scheduler, and a Dask worker container (figure \ref{fig1}).  The Dask worker container provides immediate execution resources for user tasks, providing the user with the feel of near-instant responsiveness of the system \cite{adamec2021coffea}.

Fermilab's Elastic Analysis Facility uses OKD \cite{okd} as the container orchestration engine. As mentioned above, increased security and network isolation provided by OKD are key for guaranteeing compliance with both Laboratory and DOE policies. The project started in 2018 as an effort to create a dynamic, user-oriented analysis facility capable of elastic resource allocation and dynamic scaling. The facility was designed to be a multi-VO service in preparation for high capacity data analysis demands for CMS with HL-LHC and other upcoming experiments with similar demands such as DUNE.

As a common goal, both facilities incorporate elasticity into their design: they can dispatch work not only to resources within the facility, but also to an external co-located computing platform (e.g.\ university and laboratory batch systems).  They dynamically provision worker instances using the high throughput computing scheduler HTCondor \cite{HTCondor}.

For Coffea-casa, Kubernetes not only allows programmatic service deployment but provides ``infrastructure-as-code'' where the entire infrastructure can be expressed as static (YAML) files that can be shared and used by others.  The facility utilizes a GitOps philosophy, keeping all files in a source code repository and allowing an agent (Flux) to manage synchronization between the repository and the Kubernetes cluster.  Multiple agents can deploy from the same repository, allowing separate production and development instances.  Finally, having infrastructure-as-code  allowed easily packages the core infrastructure (e.g., removing the site-specific passwords and secret keys) as a Helm chart, a format used by the Kubernetes community for application packaging \cite{helm}.

\section{Providing Reproducible, Interactive Computing Environments}

\subsection{Container images}

The Coffea-casa analysis facility at Nebraska was initially conceived to support the Coffea analysis framework \cite{DBLP:journals/corr/abs-2008-12712}.  Coffea-casa builds and uses custom containers to facilitate the integration of such a complex application.  Although the team builds and maintains these images, the versatility of Kubernetes allow for the drop-in replacement of other custom Jupyter notebook containers, and the Jupyterhub instance can be configured to allow user selection of supported images.

The Coffea-casa container also integrates a custom Coffea-casa extension to the Dask JobQueue class \cite{Dask,jobqueue}.  The JobQueue API allows for Dask tasks to be easily deployed into a batch system such as Slurm or HTCondor. The customization made by the Coffea-casa team allow for site specific parameters, such as queue name and submitter host, to be easily obscured from the user thereby making utilization of batch resources transparent to end users. 

In a similar way, Fermilab's Elastic Facility (EAF) uses containers as means to provide reproducible and reliable environments for multiple users, experiments and diverse data science requirements. However, image structuring is done in a layered pattern where ``base'' images are curated by the project with minimal software and common configurations. Later on, more complex, experiment-specific images are built from said common base, allowing for the use of caches at build, creating lightweight images, tailored for each and every use case within our user community. GPU images with specific CUDA versions and TensorFlow distributions are also derived from common bases, guaranteeing that all environment flavors contain the same (or very similar) analysis libraries and tools.

\subsection{Binderhub}

Binderhub is a Kubernetes-based cloud service that can launch a repository of code (from GitHub, GitLab, and others) in a browser window such that the code can be executed and interacted with in the form of a Jupyter notebook. Binder, the product behind mybinder.org \cite{binder} as a user interface, is also useful for reproducibility because the code needs to be version controlled and the computational environment needs to be documented in order to benefit from the functionality of Binder \cite{the_turing_way_community_2021_5671094}.

When first deploying BinderHub on Fermilab EAF, OKD's security features blocked a number of operations, including pod deletion permissions and no service accounts for performing API calls. Given that BinderHub deploys arbitrary user code on our clusters, it needs to be carefully monitored to avoid misuse or potentially malicious users to take advantage of the resource. The Binder code currently deployed on EAF has been modified at the source to ensure proper limitations on API calls, use and specification of Service Accounts, SELinux options and other key elements. The patches are in place to protect the underlying infrastructure and have been contributed upstream to BinderHub developers.

The EAF BinderHub instance is also configured as a JupyterHub service which in turn allows secure calls to JupyterHub's oauth APIs, guaranteeing consistency across applications with a central source for user data, which, in the EAF case is the Laboratory's LDAP server. Binderhub integration is planned to be delivered as a part of Coffea-casa AF deployment and will be tested in Q2 2022.

\subsection{Helm Charts}

The analysis facility model adopted by both the Fermilab EAF and Coffea-casa seeks to be modular and exportable.  To this end, each facility writes Helm charts for the various modules they write.  These modifications can then be incorporated into any specific deployment and persisted for history or any rollbacks required. Helm charts and declarative configuration in general has been key for building the EAF and ensuring high availability and reliability of services. Deploying applications in OKD using Helm and Openshift Templates allows for faster iterations and reduces operational load when dealing with cluster downtimes and other unforeseen events damaging or disabling the underlying infrastructure. Even though the EAF ecosystem is complex, it takes almost no time and effort to re-deploy it completely and re-establish service using any available OKD installation.

\subsection{GitOps}

GitOps is defined as a model for operating Kubernetes clusters or cloud-native applications using the Version Control System Git as the single source of truth \cite{beetz2021gitops}. One of the features GitOps envisions declarative descriptions of an environment to be stored as infrastructure-as-code in a Git repository.

Coffea-casa is deployed to the Nebraska Kubernetes resource via Flux \cite{flux}. Flux allows for site-specific configurations to be stored and managed in git so that administration of the analysis facility can be handled via a collaborative group of administrators and in a deterministic manner.  Flux allows for a portion of the site's configuration to be public and shareable while portions of the deployment can be retained as private. A similar deployment is in development at the Fermilab EAF, using ArgoCD \cite{argocd}.

\subsubsection{Repeatability, Reliability, and Scalability through GitOps}

Continuous integration, delivery, and deployment are software development industry practices that enable organizations to frequently and reliably release new features and products {cicdreview}. They allow for rapid collaboration, better quality control, and automation as code goes through formal review processes and is audited and tested on a regular basis.

The GitOps framework recommends managing code changes via merge/pull requests in code management systems, and implementing continuous practices for faster deployments. Both analysis facilities have implemented peer-reviewed code reviews as well as automated services for identifying new library versions and building releases, running dependency checks, and auditing images for known vulnerabilities.

An important benefit of GitOps is faster error recovery and reduced chances for failure modes. When configurations and application states are kept in tracked files, the repository becomes a history of all changes and deployment states, acting in a similar way to audit and transaction logs \cite{gitops_acm}. GitOps tools such as ArgoCD and Flux used by both analysis facilities provide ways to quickly perform deployments and rollbacks when pods become unhealthy, applications fail, or infrastructure failures occur. This reduction in the mean-time-to-recovery represents an operational advantage and allows tasks which traditionally take many hours, manpower and specific expertise, to be done in a few minutes and with lower operational risk.

\section{Summary}

The analysis facilities at Nebraska and Fermi National Laboratory represent initial successes of leveraging on-premises computing and cloud computing platform management to fundamentally change the user experience with large scale scientific computing.  The facilities have both demonstrated using Kubernetes to provide a platform for the scalable utilzation of computing resources from a customized Juypter environment.  Further development and refinement of the packaging will deliver a portable platform that could be reasonably deployed at many other sites.

\bibliographystyle{unsrtnat}
\bibliography{sample}

\begin{thebibliography}{13}
\providecommand{\natexlab}[1]{#1}
\providecommand{\url}[1]{\texttt{#1}}
\expandafter\ifx\csname urlstyle\endcsname\relax
  \providecommand{\doi}[1]{doi: #1}\else
  \providecommand{\doi}{doi: \begingroup \urlstyle{rm}\Url}\fi

\bibitem[Adamec et~al.(2021)Adamec, Attebury, Bloom, Bockelman, Lundstedt,
  Shadura, and Thiltges]{adamec2021coffea}
Matous Adamec, Garhan Attebury, Kenneth Bloom, Brian Bockelman, Carl Lundstedt,
  Oksana Shadura, and John Thiltges.
\newblock Coffea-casa: an analysis facility prototype.
\newblock In \emph{EPJ Web of Conferences}, volume 251, page 02061. EDP
  Sciences, 2021.

\bibitem[Inc.(2018)]{okd}
Red~Hat Inc.
\newblock Okd.io, 2018.
\newblock URL \url{https://www.okd.io/about/}.

\bibitem[Thain et~al.(2005)Thain, Tannebaum, and Livny]{HTCondor}
Douglas Thain, Todd Tannebaum, and Miron Livny.
\newblock Distributed computing in practice: The condor experience.
\newblock \emph{Concurrency and Computation: Practice and Experience},
  17\penalty0 (2-4):\penalty0 323--356, 2005.
\newblock \doi{http://dx.doi.org/10.5281/zenodo.2579447}.

\bibitem[for~the Cloud Native Computing~Foundation(2022)]{helm}
Helm~Authors for~the Cloud Native Computing~Foundation.
\newblock Helm, 2022.
\newblock URL \url{https://helm.sh/}.

\bibitem[Smith et~al.(2020)Smith, Gray, Cremonesi, Jayatilaka, Gutsche, Hall,
  Pedro, Flechas, Melo, Belforte, and
  Pivarski]{DBLP:journals/corr/abs-2008-12712}
Nicholas Smith, Lindsey Gray, Matteo Cremonesi, Bo~Jayatilaka, Oliver Gutsche,
  Allison Hall, Kevin Pedro, Maria~Acosta Flechas, Andrew Melo, Stefano
  Belforte, and Jim Pivarski.
\newblock Coffea - columnar object framework for effective analysis.
\newblock \emph{CoRR}, abs/2008.12712, 2020.
\newblock URL \url{https://arxiv.org/abs/2008.12712}.

\bibitem[{Dask Development Team}(2016)]{Dask}
{Dask Development Team}.
\newblock \emph{Dask: Library for dynamic task scheduling}, 2016.
\newblock URL \url{https://dask.org}.

\bibitem[jobqueue Development~Team(2018)]{jobqueue}
Dask jobqueue Development~Team.
\newblock Dask-jobqueue, 2018.
\newblock URL \url{http://jobqueue.dask.org/en/latest/}.

\bibitem[bin(2022)]{binder}
The binder project, 2022.
\newblock URL \url{https://mybinder.org/}.

\bibitem[Community(2021)]{the_turing_way_community_2021_5671094}
The Turing~Way Community.
\newblock {The Turing Way: A handbook for reproducible, ethical and
  collaborative research}, November 2021.
\newblock URL \url{https://doi.org/10.5281/zenodo.5671094}.

\bibitem[Beetz and Harrer(2021)]{beetz2021gitops}
Florian Beetz and Simon Harrer.
\newblock Gitops: The evolution of devops?
\newblock \emph{IEEE Software}, 2021.

\bibitem[Foundation(2022)]{flux}
Cloud Native~Computing Foundation.
\newblock Flux, 2022.
\newblock URL \url{https://fluxcd.io/}.

\bibitem[Project(2022)]{argocd}
Argo Project.
\newblock Argo cd, 2022.
\newblock URL \url{https://argoproc.github.io/cd}.

\bibitem[Limoncelli(2018)]{gitops_acm}
Thomas~A. Limoncelli.
\newblock Gitops: A path to more self-service it: Iac pr = gitops.
\newblock \emph{Queue}, 16\penalty0 (3):\penalty0 13–26, jun 2018.
\newblock ISSN 1542-7730.
\newblock \doi{10.1145/3236386.3237207}.
\newblock URL \url{https://doi.org/10.1145/3236386.3237207}.

\end{thebibliography}
\end{document}